# EG WEIGHTED DISTRICTS

RAY J WALLIN[*]

JANUARY 1, 2017

corrections/feedback welcome: rayjwallin01@gmail.com

[*] Ray J Wallin has been active in local politics in Saint Paul and Minneapolis, MN, writing and providing research to local campaigns. He has an engineering degree and an MBA in marketing.

# ABSTRACT

Somehow, Arxiv continues to flag this paper in Google searches as my WDM paper. (I will not expand the acronym here so it will not show up in additional searches.) This paper is a tiny section of an old working model and is of little value. For the full WDM paper, please see:

https://papers.ssrn.com/sol3/papers.cfm?abstract_id=3308888

Thank You.

This past decade has seen a noticeable uptick in asymmetric election results along with the inevitable claims of gerrymandering and litigation. Research, too, has followed, giving rise to intense scrutiny of elections, where the goal is to understand not only what goes into gerrymandering, but how to measure what comes out.

Perhaps the most cited symmetry measure of this decade is the EG. The EG has been commonplace in gerrymandering litigation nationwide and the focus of numerous articles, both popular and scholarly. This article shows how the EG can be represented as a weighting function.

INTRODUCTION

Our courts have long expressed a desire for a consistent, accurate gerrymandering measure. Deprived of a benchmark, guidance from the courts is reduced to individual rulings which are too often viewed by the public as partisan no matter the verdict. These case-by-case decisions bring the possibility of tamping down gerrymandering activity in a whack-a-mole fashion, but lack a broader interpretation to be systematically applied in lower court rulings or in the redistricting process as a line not to be crossed.

When the judicial branch dictates the actions of the legislative branch, lines must be clear, which is likely the reason the Supreme Court continues to insist on a viable standard to be applied across the board.

In the 1986 *Davis v. Bandemer* opinion, the Supreme Court agreed that partisan gerrymandering claims are justiciable yet found no standard to rule by.[1] Two decades later, the *Vieth v. Jubelirer* decision of 2004 and *LULAC* in 2006 effectively reaffirmed the mantra,[2] that the Court remains open to a standard but has found none.

The restraint exhibited in *Vieth* and *LULAC* relaxed judicial pressure on redistricters to draw fair maps which may have led to the rise in unbalanced redistricting plans following the 2010 census. But whatever the reason, this decade's lopsided election results produced an unceasing wave of public outcry with litigation following fast on its heels.

The few cases ruled on by the Supreme Court in the summer of 2018 were hoped by the gerrymandering reform community to bring verdicts of gerrymandering along with new measurement standards. What actually happened was foreseen by few. The effects of the Supreme Court's actions were negligible as they essentially handed the cases, and the problem of gerrymandering, back to the lower courts.

This article presents the EG weighting function, breaking apart the EG equations to plot them and evaluate a plan via a partisan symmetry measure which evaluates each district.

he goal of a gerrymander is to cluster one party's districts in the sweet-spot where seats are relatively safe from flipping though not yet packed with their own votes. Packing a few districts with opposition votes frees up more gerrymandering party votes in surrounding districts, ideally giving them more safe-seats.

---

[1] Davis v. Bandemer, 478 U.S. (1986)
[2] Vieth v. Jubelirer, 541 U.S. (2004); League of United Latin Am. Citizens v. Perry (LULAC), 548 U.S. (2006)

## I. THE EG WEIGHTING FUNCTION

**Table 1**
WDM and EG results for a Hypothetical Four-District State, Unshifted Data
$D_{dist}$ is the Democratic district vote-share and $P_w$ is the percentage of WV

| District | Votes | | | | Wasted Votes | | | | WDM |
| | Party | | | | Party | | | | |
| | D | R | Total | $D_{dist}$ | D | R | Total | $P_w$ | Weight |
|---|---|---|---|---|---|---|---|---|---|
| 1 | 39 | 61 | 100 | 39% | 39 | 11 | -28 | -28% | -24 |
| 2 | 49 | 51 | 100 | 49% | 49 | 1 | -48 | -48% | -5 |
| 3 | 59 | 41 | 100 | 59% | 9 | 41 | 32 | 32% | 25 |
| 4 | 69 | 31 | 100 | 69% | 19 | 31 | 12 | 12% | 12 |
| | 216 | 184 | 400 | | 116 | 84 | -32 | | |
| | | | | V = 54.0% | | | EG = -8.0% | | W = 2.0 |

**Table 2**
WDM and EG results for a Hypothetical Four-District State
after a uniform vote-shift of 2 points

| District | Votes | | | | Wasted Votes | | | | WDM |
| | Party | | | | Party | | | | |
| | D | R | Total | $D_{dist}$ | D | R | Total | $P_w$ | Weight |
|---|---|---|---|---|---|---|---|---|---|
| 1 | 41 | 59 | 100 | 41% | 41 | 9 | -32 | -32% | -25 |
| 2 | 51 | 49 | 100 | 51% | 1 | 49 | 48 | 48% | 5 |
| 3 | 61 | 39 | 100 | 61% | 11 | 39 | 28 | 28% | 24 |
| 4 | 71 | 29 | 100 | 71% | 21 | 29 | 8 | 8% | 8 |
| | 224 | 176 | 400 | | 74 | 126 | 52 | | |
| | | | | V = 56.0% | | | EG = 13.0% | | W = 3.0 |

The EG method was introduced in 2014 and has featured prominently in this decade's volley of court cases. In the court's opinion of the *League of Women Voters v. The Commonwealth of Pennsylvania*,[3] the EG is mentioned over 40 times. In the opinion of *League of Women Voters v. Rucho* in North Carolina, the EG is mentioned over 80 times.[4] In Wisconsin's *Whitford v. Gill*, the lower court opinion mentions the EG over 200 times while the SCOTUS *Whitford* opinion tones this down to just over a dozen.[5]

This section shows how the EG, which is typically presented as the percentage of WV in a state, is better understood as a weighting function. That is, the EG weights each district according to its Democratic vote-share, $D_{dist}$.

Table 1 represents a four-district hypothetical state that elected 2 Democrats and 2 Republicans. The Democratic statewide vote-share is V = 54%. From there, two votes from each Republican candidate are shifted to the Democratic candidate to give Table 2 where the statewide Democratic vote increases to V = 56% and 3 Democrats are elected.

---

[3] League of Women Voters of Pa. v. Commonwealth, 175 (Pa. 2018)
[4] Common Cause v. Rucho, 279 (M.D.N.C. 2018)
[5] District Court for the Western District of Wisconsin Remedial Opinion, Whitford v. Gill (2016); Gill v Whitford, 585 U.S. (2018)

A district's total number of WV is calculated via either Equation 1 or 2, depending on whether the Republican or Democrat candidate wins the election. The first term of each equation is the number of Republican WV and the second term is the number of Democratic WV. The difference between these terms is the total number of WV in the district.

$$Wasted\ Votes\ = (R\ votes - D\ votes)\ /\ 2\ -\ D\ votes \qquad \text{R candidate wins} \quad (1)$$

$$Wasted\ Votes\ =\ R\ votes -\ (D\ votes - R\ votes)\ /\ 2 \qquad \text{D candidate wins} \quad (2)$$

The EG is usually presented as the percentage of statewide WV in relation to the total number of statewide votes. Equivalently, the EG can be calculated as the average of $P_w$, the district percentage of WV.[6] Rewriting Equations 1 and 2 in terms of $P_w$ and $D_{dist}$, we arrive at Equations 3 and 4.

$$P_w\ =\ -50\% -\ 2\ (D_{dist} -\ 50\%) \qquad D_{dist} < 50\% \qquad (3)$$

$$P_w\ =\ \ \ 50\% -\ 2\ (D_{dist} -\ 50\%) \qquad D_{dist} > 50\% \qquad (4)$$

---

[6] Ibid, 853. McGhee at times uses district vote share (Equations 3 and 4) to calculate the EG and at other times uses the actual vote count. To eliminate turnout-bias from measurements, this article uses district vote-share; Turnout-bias arises from the use of actual votes instead of the percentage of votes and often skews results in Democrats favor by a percentage point or two. Michael D McDonald and Robin E Best, "Unfair Partisan Gerrymanders in Politics and Law: A Diagnostic Applied to Six Cases," *Election Law Journal* 14, no. 4 (December 17, 2015): 19, https://doi.org/10.1089/elj.2015.0358.; Discusses sources of turnout-bias. Michael D. McDonald, "The Arithmetic of Electoral Bias, with Applications to U.S. House Elections," *APSA 2009 Toronto Meeting Paper*, 2009.; Uses actual votes in calculations, acknowledging they introduce slight turnout-bias. John F. Nagle, "Measures of Partisan Bias for Legislating Fair Elections," *Election Law Journal: Rules, Politics, and Policy* 14, no. 4 (November 19, 2015): 346–60, https://doi.org/10.1089/elj.2015.0311.

**Figure 1**
EG Weighting Function

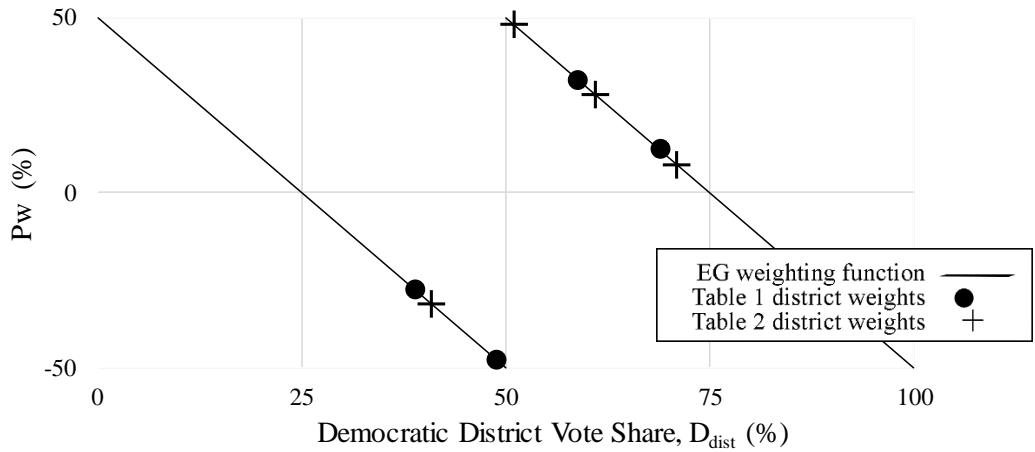

**Figure 2**
EG Response to Uniform Vote-shift

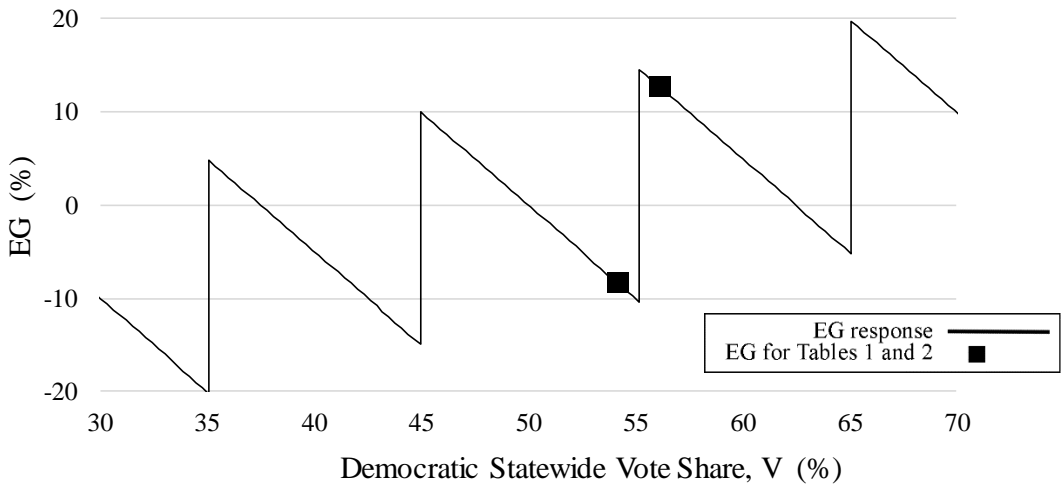

Equations 3 and 4 are defined as the EG weighting function as they assign a weight ($P_w$) to each district according to its vote-share. This weighting function, plotted in Figure 1,[7] illustrates which districts have the most WV (heavier weight) and which have the least (lighter weight).

Districts near the EG crossover points of $D_{dist} = 25\%$ and $75\%$ have near-zero WV. In contrast, closely contested elections and packed districts are weighted heavily, either positively or negatively, by the EG. In reality, few districts are so severely packed that more than, say 95% of the voters are from one party, which means the most heavily weighted districts in most plans will be near the $D_{dist} = 50\%$ discontinuity in the weighting function. Apart from the discontinuity, the slope of the weighting function is -2.

---

[7] Similar plots can be found in literature. For instance, *see*: Benjamin Plener Cover, "Quantifying Partisan Gerrymandering: An Evaluation of the EG Proposal," *Stanford Law Review* 70 (2018): 1131.

Individual district weights from Tables 1 and 2 are plotted in Figure 1 as circles and crosses, respectively. The vote-shift from Table 1 to Table 2 causes the weight to drop by 4 points in districts 1, 3, and 4, where no seat has flipped. The district 2 seat did flip, and though the magnitude of its weight did not change, the sign did, and the district jumps from nearly the lowest weight to nearly the highest, an increase of 96 points. In Table 1, EG = -8%, indicating that the state is gerrymandered to favor Republicans. In Table 2, EG = 13%, indicating that the state is gerrymandered to favor Democrats. The change in this district's weight is the major influence in the EG's jump of 21 points.

From there, a vote-shift analysis, shown in Figure 2, is performed to show EG behavior in relation to changes in voter behavior. By uniformly shifting district votes in Table 1, V is varied from 30% to 70% and the EG for each vote-share is plotted as the black sawtooth curve in Figure 2. The EG values of -8% and +13% from Tables 1 and 2 are plotted as black squares. The sawtooth behavior of the graph comes from the EG's response to a flipped seat. When a seat flips, the EG jumps 25% (one of four seats), or three times the EG gerrymandering threshold of ±7%. From one side of a jump to the other, the EG measures the plan quite differently, yet little has changed in the state, including the overall vote-share.[8]

---

[8] Discusses the erratic behavior of the EG. Mira Bernstein and Moon Duchin, "A Formula Goes to Court: Partisan Gerrymandering and the EG," *Notices of the American Mathematical Society* 64, no. 09 (October 1, 2017): 1020–24, https://doi.org/10.1090/noti1573.

## II. Discussion

Thirty years have passed since the Supreme Court expressed interest in a viable gerrymandering standard. To date, all measures presented before it have fallen short of expectations.

Politicians desire safe-seats. A gerrymander creates an abundance of safe-seats by siphoning party votes from opposition districts. The EG can be used as a weighting function. Plotting individual district weights as a function of vote-share allows the user and the courts a view of an entire election cycle in one picture.

Our courts look for durable gerrymanders. A measure of gerrymandering must be explicit, comprehensible, and produce stable results.